\documentclass[review]{elsarticle}

\usepackage{hyperref}
\usepackage{xcolor}
\usepackage{placeins}
\journal{Journal of pre-prints}

\definecolor{myColor}{rgb}{0,0,0}

\begin{document}

\begin{frontmatter}

\title{Exploring the potential of transfer learning for metamodels of heterogeneous material deformation} 

\author{E Lejeune \corref{mycorrespondingauthor}}
\author{B Zhao }
\cortext[mycorrespondingauthor]{Corresponding author: \texttt{elejeune@bu.edu}}
\address{Department of Mechanical Engineering, Boston University, Boston, MA 02215}

\begin{abstract}

From the nano-scale to the macro-scale, biological tissue is spatially heterogeneous. Even when tissue behavior is well understood, the exact subject specific spatial distribution of material properties is often unknown. And, when developing computational models of biological tissue, it is usually prohibitively computationally expensive to simulate every plausible spatial distribution of material properties for each problem of interest. Therefore, one of the major challenges in developing accurate computational models of biological tissue is capturing the potential effects of this spatial heterogeneity. Recently, machine learning based metamodels have gained popularity as a computationally tractable way to overcome this problem because they can make predictions based on a limited number of direct simulation runs. These metamodels are promising, but they often still require a high number of direct simulations to achieve an acceptable performance. Here we show that transfer learning, a strategy where knowledge gained while solving one problem is transferred to solving a different but related problem, can help overcome this limitation. Critically, transfer learning can be used to leverage both low-fidelity simulation data and simulation data that is the outcome of solving a different but related mechanical problem. In this paper, we extend Mechanical MNIST, our open source benchmark dataset of heterogeneous material undergoing large deformation, to include a selection of low-fidelity simulation results that require $\approx 2-4$ orders of magnitude less CPU time to run. Then, we show that transferring the knowledge stored in metamodels trained on these low-fidelity simulation results can vastly improve the performance of metamodels used to predict the results of high-fidelity simulations. In the most dramatic examples, metamodels trained on $100$ high fidelity simulations but pre-trained on $60,000$ low-fidelity simulations achieves nearly the same test error as metamodels trained on $60,000$ high-fidelity simulations ($1-1.5\%$ mean absolute percent error). {\color{myColor} In addition, we show that transfer learning is an effective method for leveraging data from different load cases, and for leveraging low-fidelity two-dimensional simulations to predict the outcomes of high-fidelity three-dimensional simulations.} Looking forward, we anticipate that transfer learning will enable us to better capture the influence of tissue spatial heterogeneity on the mechanical behavior of biological materials across multiple different domains. 

\end{abstract}


\end{frontmatter}


\section{Introduction}
\label{sec:intro} 

{\color{myColor} Mechanical models of biological materials are useful for applications ranging from understanding the role of mechanics during development to predicting the outcomes of possible medical interventions \citep{ambrosi2011perspectives,lejeune2016tri}.}
And, the complex heterogeneous behavior of biological materials makes constructing relevant mechanical models a both challenging and interesting problem \citep{genet2015heterogeneous}. 
Biological materials typically have a spatially heterogeneous micro- and nano- structure \citep{gu2017printing,lejeune2019analyzing}. In addition, the material properties of biological tissue often vary on macroscopic length scales \citep{bersi2020multimodality}.  
Computational modeling is the ideal tool for directly investigating the effects of this heterogeneity \citep{hugenberg2020characterization,yu2019artificial}. 
With computational modeling, it is possible to capture spatial heterogeneity and subsequently measure its effects \citep{Gokhale+etal08}. 
Critically, making meaningful predictions about the behavior of biological materials will often require multiple simulations to account for both subject-specific variations and model uncertainty \citep{Lee+etal18,tran2019uncertainty}. 
In other words, rather than representing a biological material through the outcome of a single simulation, examining the outcomes of a suite of simulations is often more appropriate \citep{Lejeune+etal19b,Lejeune+etal19}. 
However, particularly when simulating complex microstructure and/or whole organ geometries, this approach can be prohibitively computationally expensive \citep{yang2010constrained}. 
For example, it is usually prohibitively computationally expensive to simulate every plausible spatial distribution of material properties for a given problem of interest \citep{Bessa+etal17,gu2018novo,liu2016self}.

\begin{figure}[h]
\begin{center}
\includegraphics[width=.4\textwidth]{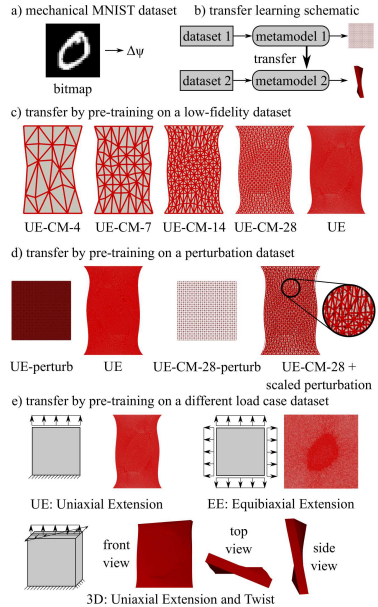}
\caption{\label{fig:intro} a) The ``Mechanical MNIST'' dataset contains FEA simulation results, in this paper we consider the relationship between the input bitmap that prescribes material properties and the change in strain energy due to applied displacement; b) A schematic of a model trained on one dataset being transferred to make predictions on another dataset; c) Illustration of the levels of mesh refinement explored in our low fidelity models; d) An illustration of a perturbation compared to the final level of applied displacement, we note that the scaled result of the perturbation is an imperfect match to the final deformed configuration; e) Schematic illustration of the uniaxial extension (UE), equibiaxial extension (EE), {\color{myColor}and three-dimensional uniaxial extension and twist (3D)} load cases.}
\end{center}
\end{figure}

One way to overcome this limitation is through constructing metamodels \citep{Hanakata+etal18}. Metamodels, ``models of models'' often referred to as ``surrogate models,'' are used to predict model outputs -- often referred to as quantities of interest (QoI) -- from given model input parameters \citep{Lejeune20b,Queipo+etal05}. 
For example, in the work that we present here, we focus on predicting the QoI change in strain energy, $\Delta \psi$, from a bitmap that defines a spatial distribution of material properties. This is illustrated in Fig. \ref{fig:intro}a. 
Metamodels are constructed (i.e. trained) from the outcomes of full-fidelity model runs \citep{Wang+Sun19}. 
Typically, metamodels are constructed to be computationally cheap to execute \citep{Vu+etal17,Yang+Perdikaris19}. 
For example, a trained metamodel will typically execute in fractions of a second while the equivalent finite element simulations typically take orders of magnitude longer to complete \citep{Wang+Sun18}. 
If a metamodel is successfully trained, it will be able to make accurate predictions for input parameter combinations outside of those used to generate training simulations and thus enable a computationally cheap exploration of the input parameter space \citep{Teichert+Garikipati+etal19a}. 
This means that metamodels are an invaluable tool for conducting design optimization, sensitivity analysis, uncertainty quantification, and some forms of multiscale modeling \citep{Alber+etal19,Bessa+etal19,Peirlinck+etal19,Sahli+etal18}. 

For metamodels that predict the mechanical behavior of heterogeneous material, convolutional neural networks (CNNs) are a popular choice \citep{Schwarzer+etal19,yang2019using,zhang2020finite}.
{\color{myColor} This is because CNNs are designed to capture spatial relationships \citep{yang2020prediction}.
Recently, CNNs have been used to capture the mechanical behavior of heterogeneous materials with applications such as fracture prediction \citep{hsu2020using}, design optimization \citep{yu2019artificial}, and material homogenization \citep{rao2020three}.}
However, CNNs are notorious for requiring a large amount of training data \citep{d2020structural}.
This means that generating sufficient training data will still be prohibitively computationally expensive in many cases. 
Transfer learning, where knowledge gained while solving one problem is transferred to a different but related problem, may help us overcome this issue \citep{Pan+etal09}.
Specifically, if easy to acquire data (either computationally cheap or generated for a previous study) can be leveraged, far fewer harder to acquire or new simulations will be necessary. 
A schematic of transfer learning is illustrated in Fig. \ref{fig:intro}b. 
In the context of simulating biological tissue with the finite element method \citep{Costabal+etal19,Javili+etal15}, examples of easier to acquire simulation data include simulations conducted with linear elements where quadratic elements would be more appropriate, simulations conducted on an un-refined mesh, simulations with less complexity, simulations in a lower dimensional space, and simulations of an identical tissue run previously to solve a different problem.
{\color{myColor}We note that recently multi-fidelity modeling, a specific form of transfer learning where low fidelity simulation data is leveraged to make predictions about high fidelity simulation data, has gained popularity for constructing metamodels of computationally expensive simulations \citep{Bonfiglio+etal18}. }
Here we choose a few relevant examples of simulation data from our Mechanical MNIST benchmark dataset (details provided in Section \ref{sec:dataset}) to investigate the efficacy of metamodeling via transfer learning for problems related to heterogeneous tissue undergoing large deformation.  
Examples of the ``different but related'' datasets that we investigate in this paper are shown in Fig. \ref{fig:intro}c-e. 

In this paper, we begin in Section \ref{sec:methods} by describing our dataset, our metamodeling strategy, and our chosen method for transfer learning. Then, in Section \ref{sec:res}, we describe and discuss the results of our study. Critically, we find that transfer learning is an effective way to leverage data from low-fidelity simulations, perturbation simulations, and simulations with different loading conditions. Finally, we conclude in Section \ref{sec:conclusion}. In Section \ref{sec:additional_info}, we provide information on how to access all of the data and the code required to reproduce our results. 

\section{Methods}
\label{sec:methods}

Here we begin in Section \ref{sec:dataset} with a brief description of the Mechanical MNIST dataset and the additions to the dataset that were required to conduct the research presented in this paper \cite{Lejeune20a}. Then, in Section \ref{sec:metamodel}, we briefly describe our metamodel selection and in Section \ref{sec:tl} we describe our methods for implementing transfer learning. We end in Section \ref{sec:mf_note} with a brief note on alternative approaches to transfer learning and multi-fidelity modeling. Details for accessing all of the relevant data and software are provided in Section \ref{sec:additional_info}. 

\subsection{Mechanical MNIST dataset}
\label{sec:dataset}

The Mechanical MNIST dataset \citep{Lejeune20a} is a benchmark dataset with data from finite element simulations inspired by the MNIST dataset of handwritten digits popular within the computer vision research community \citep{LeCun+etal98}. 
Briefly, the dataset is constructed by using the MNIST bitmaps, illustrated in Fig. \ref{fig:intro}, to define the material properties of a heterogeneous block of Neo-Hookean material. The white pixels correspond to Young's Modulus $E=100$, the black pixels correspond to $E=1$, and $E$ for the grayscale pixels is computed by linearly interpolating. 
The two-dimensional block of material is then subject to an applied displacement, where boundary conditions for the Uniaxial Extension (UE), Equibiaxial Extension (EE) {\color{myColor} and three-dimensional extension and twist (3D)} load cases are shown schematically in Fig. \ref{fig:intro}e. 
The total change in strain energy $\Delta \psi$ is computed for each level of applied displacement. As illustrated in Fig. \ref{fig:intro}a, we consider a single QoI, $\Delta \psi$, and train our metamodels to predict this QoI when given an input bitmap. 
Identical to the original MNIST dataset, each subset of Mechanical MNIST contains $60,000$ training data points and $10,000$ test data points. 
Further details are presented in the manuscript accompanying the original dataset \citep{Lejeune20a} and in \ref{apx:simulation_info}. 
Despite the ``toy problem'' nature of the dataset, it is specifically focused on heterogeneous materials undergoing large deformation and is thus relevant to our broader understanding of the behavior of heterogeneous soft tissue. 

\begin{table}[h]
{\footnotesize
\begin{tabular}{cp{8cm}}
\hline
Name             & Description                                                                                                                            \\ \hline
UE               & uniaxial extension, full fidelity dataset (fully refined mesh, quadratic triangular elements, applied displacement is $1/2$ of a side length) \\
EE               & equibiaxial extension, full fidelity dataset                                                                                           \\
{\color{myColor} 3D} & uniaxial extension and out of plane twist, full fidelity three dimensional dataset (fully refined mesh, quadratic tetrahedral elements, applied displacement is $1/7$ of a side length, twist is $\pi/8$ radians, block thickness is $1/7$ of a side length) \\
UE-CM-28         & uniaxial extension, $28 \times 28 \times 2$ linear triangular elements                                                                            \\
UE-CM-14         & uniaxial extension, $14 \times 14 \times 2$ linear triangular elements                                                                            \\
UE-CM-7          & uniaxial extension, $7 \times 7 \times 2$ linear triangular elements                                                                              \\
UE-CM-4          & uniaxial extension,  $4 \times 4 \times 2$ linear triangular elements                                                                             \\
UE-perturb       & uniaxial extension, applied displacement is a perturbation                                                                             \\
UE-CM-28-perturb & uniaxial extension, $28 \times 28 \times 2$ linear triangular elements, applied displacement is a perturbation                                    \\ \hline
\end{tabular}}
\caption{\label{tab:datasets} Summary of the different datasets, all within Mechanical MNIST, that we investigate in this paper. {\color{myColor} The perturbation displacement is $0.001$ units compared to the side length of $28$ units. Two additional datasets, UE-CM-7-quad and UE-CM-4-quad, are described in \ref{apx:other_data}.} All data is freely available with access details provided in Section \ref{sec:additional_info}.}
\end{table}

In this paper, we use {\color{myColor}nine} different datasets that all fall under the broader Mechanical MNIST umbrella. We provide a summary in Table \ref{tab:datasets}. For this paper specifically, we created the four coarse mesh (``CM'') datasets. Other than changing the mesh size and switching from quadratic to linear elements, these simulations are identical to the simulations in the ``UE'' high-fidelity dataset. However, because the MNIST input bitmaps are $28 \times 28$, the grids coarser than $28 \times 28$ are unable to perfectly resolve the spatial pattern. {\color{myColor} In addition we generated the three-dimensional dataset ``3D'' specifically for this paper.} All other datasets had already been generated and published \citep{MMequi,MMuniaxial}. The perturbation datasets ``UE-perturb'' and ``UE-CM-28-perturb'' are contained within ``UE'' and ``UE-CM-28'' respectively. {\color{myColor}We also detail two additional datasets ``UE-CM-7-quad'' and ``UE-CM-4-quad'' in \ref{apx:other_data}.} The open source finite element library FEniCS was used to run all simulations \citep{Alnaes+etal15,Logg+etal12}. 

\subsection{Metamodel}
\label{sec:metamodel}

For each dataset shown in Table \ref{tab:datasets}, we train a metamodel to predict $\Delta \psi$ from the input bitmap.
In all cases, we used the same metamodel architecture. 
Specifically, we use a convolutional neural network (CNN) \citep{LeCun+etal15} with a convolutional layer with $20$ feature maps, followed by a max pooling layer, followed by a convolutional layer with $50$ feature maps, followed by a max pooling layer, followed by a fully connected layer with $50$ nodes. The output layer is a single node that predicts $\Delta \psi$. We use rectified linear unit (ReLU) activation functions \citep{ramachandran2017searching}. 
We construct and train the CNNs with open source library PyTorch \citep{Paszke+etal17}. 
Details for accessing the code used to implement our metamodel are given in Section \ref{sec:additional_info}. 
Metamodel performance is discussed in Section \ref{sec:mm_res}. 
{\color{myColor} We note that our investigation of potential metamodels was not exhaustive. This model architecture is a slight modification of our previously published baseline model that was chosen based on model architectures that achieves good performance on the original MNIST dataset \citep{Lejeune20a}. To our knowledge, the performance that we achieve in this paper represents the new best performance on the Mechanical MNIST datasets. However, we anticipate that this performance will be improved upon in the future.}  

\subsection{Transfer learning}
\label{sec:tl}

\begin{figure}[h]
\includegraphics[width=\textwidth]{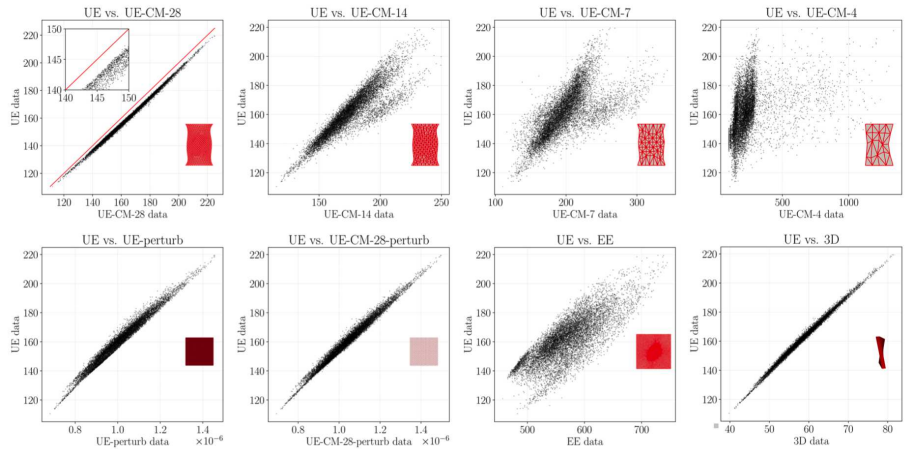}
\caption{\label{fig:raw_data} All plots illustrate the correlation between the standard Uniaxial Extension (UE) dataset and another dataset: UE-CM-28, UE-CM-14, UE-CM-7, UE-CM-4, UE-perturb, UE-CM-28-perturb, EE, {\color{myColor} and 3D}. All axis represent the QoI, change in strain energy $\Delta \psi$ at the end of the simulation, for each dataset. Each point corresponds to an input bitmap in the test set ($10,000$ points total per plot).}
\end{figure}

The essential idea behind transfer learning is that knowledge can be transferred between different but related problems. We note that making predictions based on each dataset listed in Table \ref{tab:datasets} can be thought of as a different but related problem. 
{\color{myColor} The relationship between these datasets is illustrated in Fig. \ref{fig:raw_data}, where each datapoint corresponds to one input bitmap in the Mechanical MNIST test set, and each plot shows the relationship between ``UE'' and another dataset.} In all cases, the relationship between the two datasets is clearly not entirely random.
With these datasets, we are interested in understanding the potential for transfer learning for three key applications:
\begin{itemize}
\item[1.] Leveraging low fidelity data to predict high fidelity results (UE-CM-28, UE-CM-14, UE-CM-7, UE-CM-4 $\rightarrow$ UE), {\color{myColor} UE-CM-28-perturb $\rightarrow$ 3D}) 
\item[2.] Leveraging perturbation results to predict final simulation results (UE-perturb, UE-CM-28-perturb $\rightarrow$ UE), {\color{myColor} UE-CM-28-perturb $\rightarrow$ 3D}) 
\item[3.] Leveraging one load case results to predict another load case results (EE $\rightarrow$ UE, UE $\rightarrow$ EE, {\color{myColor} UE-CM-28-perturb $\rightarrow$ 3D}) 
\end{itemize} 
{\color{myColor} The exploration of UE-CM-28-perturb $\rightarrow$ 3D is related to all three applications}. As discussed subsequently in Section \ref{sec:mf_note}, there are multiple different approaches to transfer learning. In this paper, we focus exclusively on transfer learning through model pre-training. 
With this approach, the model is ``pre-trained'' on the first dataset, and then further trained on the actual dataset of interest \citep{yosinski2014transferable}. Typically, the size of the first dataset used for pre-training will be much larger than the actual dataset of interest \citep{kawano2014automatic,yanai2015food}.  
With this method, transfer learning is worthwhile when the metamodel with initialized weights based on pre-training is able to outperform a metamodel with randomly initialized initial weights {\color{myColor} at a lower total computational cost for generating new simulations}.  
We note that in PyTorch implementing model pre-training is essentially trivial. The only additional steps beyond what is necessary to implement the metamodel described in Section \ref{sec:metamodel} are: (1) train an initial metamodel (pre-train) using the same network architecture, and (2) load the initial metamodel weights before training the actual metamodel of interest \cite{Paszke+etal17}. 

\subsection{Note on multi-fidelity modeling}
\label{sec:mf_note}
In the broader computational mechanics literature, there has been substantial recent interest in ``multi-fidelity modeling'' where, similar to what we describe in Section \ref{sec:tl}, metamodels are constructed based on simulation data sources of varying fidelity \citep{Bonfiglio+etal18}. 
For example, computationally cheap one-dimensional models have been leveraged to make predictions about three-dimensional full-fidelity models \citep{Costabal+etal19}. 
The multi-fidelity modeling paradigm has also been used to inform predictive models in the experimental setting with larger amounts of ``low-fidelity'' information obtained from simulations paired with comparatively few physical experiments \cite{lu2020extraction}. 
Multi-fidelity models are not restricted to convolutional neural networks, we note that {\color{myColor}Gaussian processes} are also popular techniques for this approach \citep{lee2020propagation}. 
Conceptually, multi-fidelity models are a form of transfer learning. 
{\color{myColor}
Notably, many multi-fidelity modeling methods presented in the literature involve specialized model architecture and/or problem specific methods for data integration \citep{zhang2019machine}. 
In this work, our goal is to explore the efficacy of transfer learning exclusively through straightforward model pre-training.
We favor this approach because it is both simple to implement, and, as demonstrated in Section \ref{sec:res}, highly effective for the problems that we are interested in.}
We note that another approach popular in the broader machine learning literature is to use a pre-trained model as a fixed feature extractor and only perform additional training on part of the model \citep{salberg2015detection}. {\color{myColor} For example, it is common to hold the weights of all layers but the final layer fixed during the second round of training, or add additional layers to the pre-trained model and only adjust the added layers to the new dataset. In the approach presented in this paper, all model weights are free to update when the new dataset is added.}
Because we have made our datasets entirely open source, it is possible for other researchers to implement alternative methods and potentially exceed the baseline performance that we report in Section \ref{sec:res}.

\section{Results and Discussion}
\label{sec:res}

In this Section, we report the results of training convolutional neural networks to predict change in strain energy $\Delta \psi$ at fixed levels of applied displacement. 
We begin in Section \ref{sec:mm_res} by showing the performance of our metamodel on each of the datasets listed in Table \ref{tab:datasets}. Then, in Section \ref{sec:pre_train_lofi}, we show the efficacy of pre-training on low-fidelity simulation, in Section \ref{sec:pre_train_perturb} we show the efficacy of pre-training on perturbation results, and in Section \ref{sec:pre_train_load} we show the efficacy of pre-training on a different load case. We end in Section \ref{sec:limit} with a brief discussion of the limitations of this study and how these limitations could be addressed in future work. 

\begin{figure}[h]
\begin{center}
\includegraphics[width=.5\textwidth]{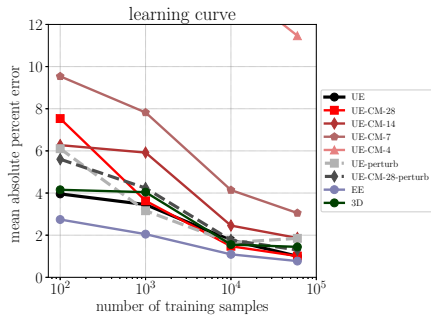}
\caption{\label{fig:learning_curve} This plot shows the test performance of each metamodel (see Section \ref{sec:metamodel}) with respect to the number of training samples for each dataset. We note that the metamodel trained on the UE dataset has $3.96\%$ mean absolute percent error when trained with $100$ samples, and $1.00\%$ error when trained with $60,000$ samples.}
\end{center}
\end{figure}

\subsection{Metamodel without pre-training}
\label{sec:mm_res}

Here we show the performance of the metamodel described in Section \ref{sec:metamodel} on the datasets listed in Table \ref{tab:datasets}. 
In Fig. \ref{fig:learning_curve}, mean absolute percent error on the test set is plotted with respect to the number of samples used to train the metamodel.
With our chosen metamodel architecture and PyTorch implementation (see Section \ref{sec:additional_info} for information on accessing the code), the UE dataset metamodel has $3.96 \%$ mean absolute percent error when trained with $100$ samples and $1.00 \%$ error when trained with $60,000$ samples. 
We note that this is an incremental improvement over previously published baseline model performance of $1.9 \%$ mean absolute percent error obtained with a CNN using $60,000$ training points \citep{Lejeune20a}. This metamodel architecture has comparable performance on all other datasets except for UE-CM-4 which, despite improved performance as more training data is added, has $37.5\%$ test error with $100$ training points and $11.5\%$ test error with $60,000$ training points. 

\begin{figure}[p]
\begin{center}
\includegraphics[width=\textwidth]{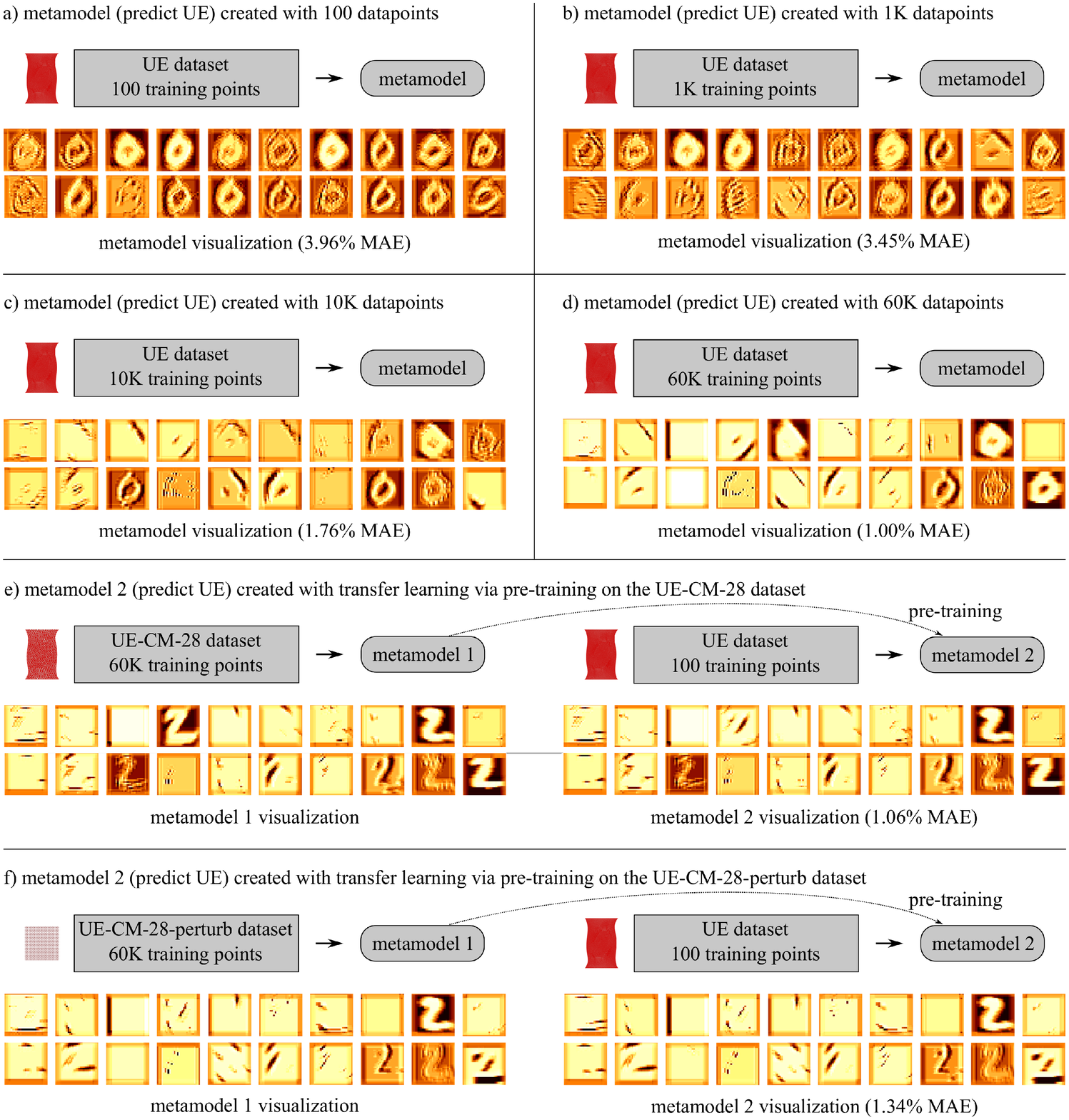}
\caption{\label{fig:vis_train}{\color{myColor} Illustration of first layer activations for an example input bitmap for different numbers of training points from the UE dataset: a) $100$ training points; b) $1,000$ training points; c) $10,000$ training points; d) $60,000$ training points. Illustration of first layer activations for an example input bitmap for pre-trained metamodels: e) a metamodel pre-trained on the UE-CM-28 dataset is then transferred to make predictions on the UE dataset; f) a metamodel pre-trained on the UE CM-28-perturb dataset is then transferred to make predictions on the UE dataset. The ``metamodel visualization'' shows the first layer activations visualized with guided backpropagation \citep{springenberg2014striving,uozbulak_pytorch_vis_2019}. Mean absolute percent error (MAE) is reported.}}
\end{center}
\end{figure}

In Fig. \ref{fig:vis_train}a-d, we show a visualization of the metamodel as more samples are used for training. Specifically, we show the first layer activations for all $20$ feature maps for a representative input bitmap using guided backpropagation \cite{springenberg2014striving}. This visualization allows us to qualitatively compare different metamodels as a supplement to our quantitative error comparisons. We note that as the number of training samples increases, the first layer activations change substantially. We will return to this qualitative comparison when we show the results of metamodel pre-training. With pre-training, the first layer activations will look much more like Fig. \ref{fig:vis_train}d, despite being trained with the same small number of points as the metamodel shown in Fig. \ref{fig:vis_train}a. 

\subsection{Metamodel pre-trained on a low-fidelity simulation dataset}
\label{sec:pre_train_lofi}

\begin{figure}[p]
\includegraphics[width=\textwidth]{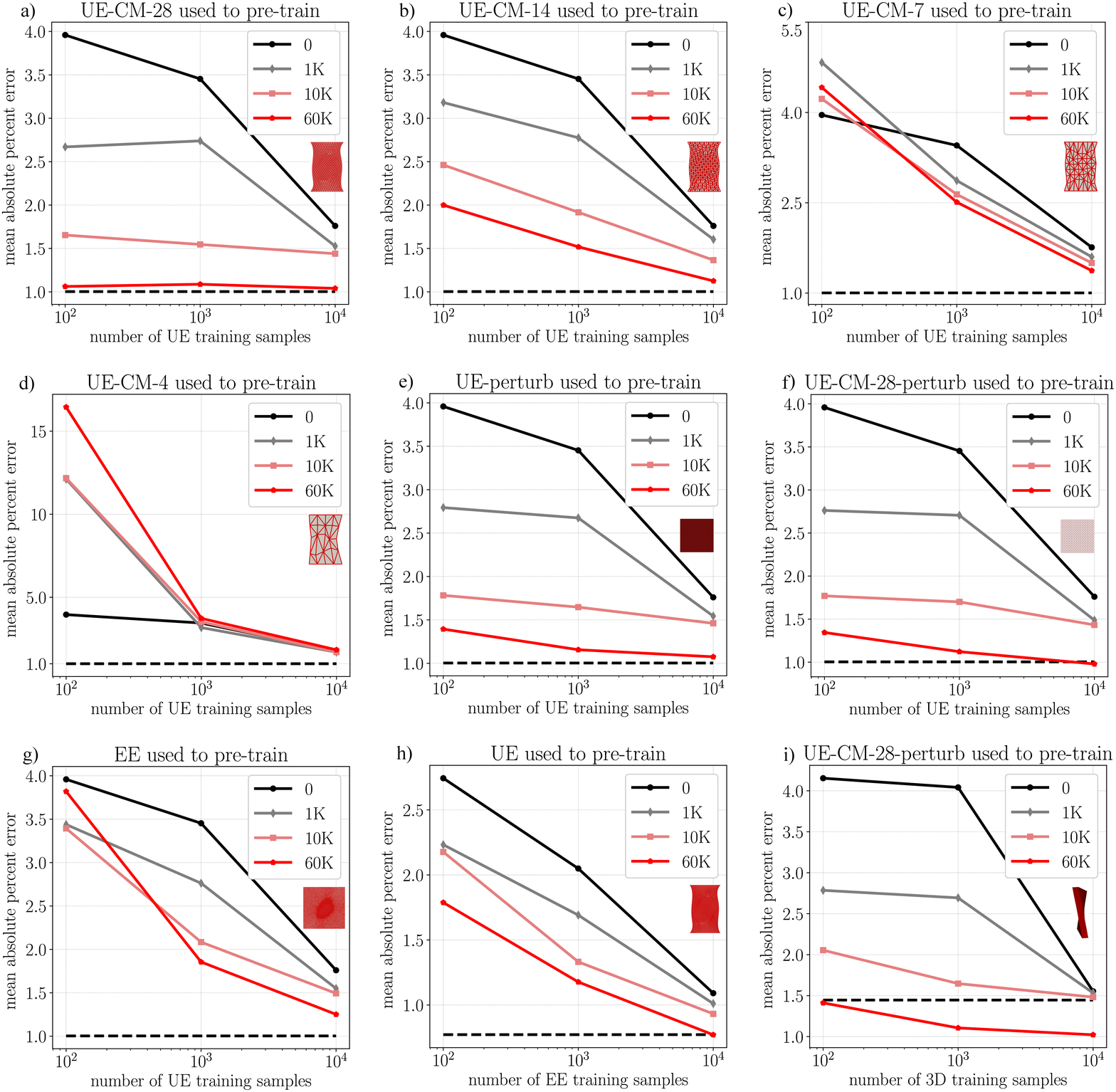}
\caption{\label{fig:plot_transfer_res} These plots shows the test performance of pre-trained metamodels with respect to the number of training samples. In all plots, the black curve labeled ``$0$'' corresponds to metamodel performance without pre-training and the black dashed line corresponds to metamodel performance without pre-training for $60,000$ training points which represents the best performance achieved by the metamodel given the available data (see Section \ref{sec:mm_res}). The other curves, labeled ``$1$K'', ``$10$K'', and ``$60$K'', correspond to pre-trained metamodel performance where each metamodel is pre-trained with the indicated number of points from the related dataset. The plots show: a) UE prediction with UE-CM-28 used for pre-training; b) UE prediction with UE-CM-14 used for pre-training; c) UE prediction with UE-CM-7 used for pre-training; d) UE prediction with UE-CM-4 used for pre-training; e) UE prediction with UE-perturb used for pre-training; f) UE prediction with UE-CM-28-perturb used for pre-training; g) UE prediction with EE used for pre-training; h) EE prediction with UE used for pre-training; {\color{myColor} i) 3D prediction with UE-CM-28-perturb used for pre-training. Note that the y-axis scaling is not the same for each plot.}}
\end{figure}

One of the most appealing applications of transfer learning for metamodel creation is the prospect of leveraging the results of computationally cheap simulations. 
In our Mechanical MNIST dataset, each of the simulations has quadratic elements and a highly refined mesh and thus takes several minutes to run on a single CPU ({\color{myColor} see \ref{apx:simulation_info} for more details}).
For many applications relevant to biomedical engineering, for example whole organ simulation, it is common for simulations to take tens to thousands of CPU hours to run \citep{rausch2017computational,terahara2020heart}. 
Changes such as switching from quadratic to linear elements and coarsening the finite element mesh can dramatically reduce computational cost, but these changes will also introduce error. For our UE-CM datasets, simulations take seconds to run on a single CPU, and yield low-fidelity predictions that, as illustrated in Fig. \ref{fig:raw_data}, are imperfect but correlated with the high-fidelity simulation results. 
Here we show what happens to our UE metamodel when it is pre-trained with our low-fidelity datasets. 

In Fig. \ref{fig:plot_transfer_res}a-d, we show metamodel test error with respect to number of training samples for the UE metamodel (black curve) and then the UE metamodel pre-trained with different numbers of training points from the UE-CM datasets. 
As illustrated in Fig. \ref{fig:raw_data}, the UE-CM-28 dataset has the closest relationship to the UE dataset while the relationship degrades as the mesh coarsens. 
In Fig. \ref{fig:plot_transfer_res}a, we see that pre-training with UE-CM-28 allows the UE pre-trained metamodel to achieve lower test error with fewer training samples. And, increasing the number of UE-CM-28 datapoints used for training improves the pre-trained metamodel performance. In Fig. \ref{fig:plot_transfer_res}b we see similar but less dramatic results for the UE-CM-14 dataset. Then, in Fig. \ref{fig:plot_transfer_res}c, we see an initially worse performance at $100$ training points for pre-training with UE-CM-7 followed by a marginal reduction in error for the UE pre-trained metamodels trained with $1,000$ and $10,000$ points. In Fig. \ref{fig:plot_transfer_res}d, we see that pre-training with the UE-CM-4 dataset is detrimental to model performance. {\color{myColor} For UE-CM-4, the combination of such an extremely coarse mesh and linear elements led to erroneous simulation results that could not be leveraged with our current method. In \ref{apx:other_data}, we provide some additional results with a coarse mesh and quadratic elements to separate the two sources of lower data quality.}

These results show that with the metamodel architectures described in this paper it is possible to use transfer learning to achieve improved performance by pre-training on low-fidelity data. Pre-training with UE-CM-28 and UE-CM-14 showed clear and consistent benefit, while pre-training with UE-CM-7 and UE-CM-4 did not. All of the datasets are freely available, thus we welcome others to implement methods that surpass this baseline performance and are potentially able to leverage the UE-CM-7 and UE-CM-4 datasets. 
{\color{myColor}In Fig. \ref{fig:vis_train}a-d, we visualize the first layer activations for the UE metamodel without pre-training for different numbers of training points, and in Fig. \ref{fig:vis_train}e we visualize both the UE-CM-28 $60,000$ training points metamodel and the subsequent pre-trained UE metamodel with $100$ high-fidelity training points. The comparison between Fig. \ref{fig:vis_train}a-d and Fig. \ref{fig:vis_train}e shows the qualitative similarity between the UE full-fidelity metamodel with a large number of training points and the pre-trained metamodel with a small number of training points.} Essentially, the metamodels are qualitatively similar and the pre-training process makes it possible to leverage this similarity.

\subsection{Metamodel pre-trained on a perturbation simulation dataset}
\label{sec:pre_train_perturb}

Building on the theme of leveraging the results of computationally cheap simulations, here we explore the potential benefits of pre-training on the results of simulated perturbations. In this case, this means simulation results from applied displacements orders of magnitude smaller than the actual applied displacement of interest. As illustrated in Fig. \ref{fig:raw_data}, both the UE-perturb and the UE-CM-28-perturb datasets are clearly related to the UE dataset. {\color{myColor} We note that in order to avoid numerical errors during the metamodel training process, we multiply the perturbation $\Delta \psi$ values by $10^8$ before training.\footnote{\color{myColor}without this step, QoIs are $\mathcal{O}(10^{-6})$ and, at the time of publication, training on the $\mathcal{O}(10^{-6})$ values in PyTorch led to a metamodel that predicted $0.0$ for all values \citep{Paszke+etal17}}} 
 In Fig. \ref{fig:plot_transfer_res}e-f, we show metamodel test error with respect to the number of training samples for the UE metamodel (black curve) and then the UE metamodel pre-trained with different numbers of training points from the UE-perturb and the UE-CM-28-perturb datasets. 
In both cases, pre-training leads to the UE pre-trained metamodel achieving lower test error with fewer training samples. And, increasing the number of datapoints used for pre-training improves the metamodel performance. In Fig. \ref{fig:vis_train}f, we also qualitatively compare the metamodel first layer activations for the UE-CM-28-perturb dataset and the UE dataset pre-trained with CM-28-perturb. 

\subsection{Metamodel pre-trained on an alternative load case dataset}
\label{sec:pre_train_load}

\begin{figure}[h]
\begin{center}
\includegraphics[width=\textwidth]{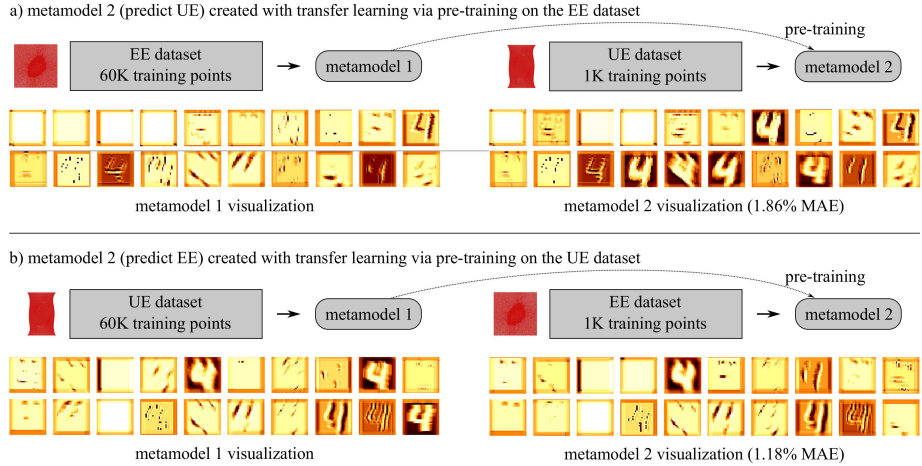}
\caption{\label{fig:vis_transfer_load} Illustration of first layer activations for an example input bitmap for pre-trained metamodels: a) a metamodel pre-trained on the EE dataset is then transferred to make predictions on the UE dataset; b) a metamodel pre-trained on the UE dataset is then transferred to make predictions on the EE dataset. The ``metamodel visualization'' shows the first layer activations visualized with guided backpropagation \cite{springenberg2014striving,uozbulak_pytorch_vis_2019}. Mean absolute percent error (MAE) is reported.}
\end{center}
\end{figure}

Switching between load cases exemplifies the transfer learning concept of a ``different but related problem.'' 
Practically, the ability to leverage data acquired for one load case and apply it to another with minimal additional simulation runs required is a powerful tool for exploring the model parameter space beyond solely the effects of heterogeneous material properties. 
In Fig. \ref{fig:raw_data}, we show that the UE and EE datasets are related, but perhaps less similar than some of the UE and UE-CM datasets. 
In Fig. \ref{fig:plot_transfer_res}g, we show metamodel test error for UE metamodels pre-trained on the EE datasets. 
In Fig. \ref{fig:plot_transfer_res}h, we show metamodel test error for EE metamodels pre-trained on the UE datasets. 
And, in Fig. \ref{fig:vis_transfer_load}, we qualitatively compare the metamodel first layer activations. 
The plots in Fig. \ref{fig:plot_transfer_res} show that pre-training the metamodel on data from another load case can be an effective way to reduce metamodel error.
And, using more points for pre-training corresponds to lower metamodel error. The only exception to this is the UE metamodel pre-trained on the EE metamodel trained with $60,000$ data points. We anticipate that it would be possible to adjust model architecture {\color{myColor} or training parameters to overcome what is likely overfitting to the newly provided data}, but we chose not to in order to maintain consistency throughout the paper. Overall, we show that metamodel pre-training for transferring between load cases is viable. 

{\color{myColor}In Fig. \ref{fig:plot_transfer_res}i, we show metamodel test error for 3D metamodels pretrained on the UE-CM-28-perturb dataset. In Fig. \ref{fig:plot_transfer_res}i, we see that a metamodel pre-trained with $60,000$ UE-CM-28-perturb datapoints and subsequently trained with $100$ 3D datapoints achieves a slightly lower test error than a metamodel trained with $60,000$ 3D datapoints. This example ties together the exploration of low fidelity datasets, perturbation datasets, and alternative load case datasets. Notably, the computational cost of generating the simulations for the transfer learning based metamodel is $\approx 0.13$\% of the computational cost for generating the $60,000$ 3D simulations, see Table \ref{tab:comp_cost} for additional details. This shows the potential for transfer learning to have an important impact in computational modeling of heterogeneous materials undergoing large deformation.}

\subsection{Limitations of the present study}
\label{sec:limit}

In our opinion, there are two main limitations to the present study. 
First, we are not able to claim that any of our metamodels are the best possible metamodels for a given dataset or that pre-training will \textit{always} improve metamodel performance. It is possible that other researchers would be able to design metamodels that vastly exceed the performance reported here. And, it is possible that other approaches that leverage different but related data would be able to exceed the performance of our model pre-training based approach with either comparable or smaller data requirements.
To address this limitation, we have released all of the data that our results are based on and welcome others to create models that exceed the baseline performance shown here \citep{MMmultifi}. This speaks to the broader problem of lack of benchmark data for method comparison within the computational mechanics research community \citep{VisionWorkshop}. 
 Second, the Mechanical MNIST dataset that we used does not represent all possible qualities of heterogeneous soft tissue. 
 For example, the current form of the dataset lacks a version with an anisotropic material model, a version with coupled fields, a version with three-dimensional simulations, and a version where the spatially heterogeneous distribution of material properties reflects specific types of biological tissue. 
To address this limitation, we plan to add to the dataset in the future and we welcome others to either build on our open source software (see Section \ref{sec:additional_info}) to generate simulations or create benchmark datasets with their own software that best address domain specific problems of interest. 
 
\section{Conclusion}
\label{sec:conclusion}
In this paper, we demonstrate that transfer learning via metamodel pre-training is a powerful tool for creating metamodels of heterogeneous material behavior. 
We demonstrate the efficacy of the approach by example using our open source Mechanical MNIST benchmark dataset. 
First, we show that convolutional neural networks are an effective tool for creating metamodels based on our datasets. 
Then, we demonstrate that pre-training on data from simulations conducted with a coarse mesh, pre-training on data from simulated perturbations, and pre-training on alternative load cases can all improve metamodel performance. 
In the most dramatic example of the efficacy of transfer learning, we show that pre-training with the UE-CM-28 (coarse mesh) and the UE-CM-28-perturb (coarse mesh perturbation simulation) datasets makes metamodels trained with $100$ high-fidelity simulation results perform almost as well as metamodels trained with $60,000$ high-fidelity simulation results for both two-dimensional and three-dimensional high fidelity simulations ($\approx 1 \%$ error and $\approx 1.5 \%$ error respectively).
In addition to this exploration of transfer learning, we have also contributed both low-fidelity and three-dimensional simulation results to the Mechanical MNIST benchmark dataset. This contribution will enable others to directly compare competing methods that may exceed the baseline performance shown here. Dataset and software access details are provided in Section \ref{sec:additional_info}. Looking forward, we anticipate that other researchers interested in creating metamodels of heterogenous materials will be able to directly implement the methods presented in this paper. 
{\color{myColor} These methods will make it computationally tractable to explore the true behavior of spatially heterogeneous soft tissue and thus better understand and predict the true mechanical behavior of biological materials.}
Overall, we anticipate that transfer learning will enable substantial future advances in soft tissue simulation for both basic research and clinical applications.

\section{Acknowledgements}

We would like to thank the staff of the Boston University Research Computing Services and the OpenBU Institutional Repository (in particular Eleni Castro) for their invaluable assistance with generating and disseminating the Mechanical MNIST datasets. This work was made possible though the Engineering Research Centers Program of the National Science Foundation (No. EEC-1647837), the Boston University Hariri Institute Junior Faculty Fellows program, and start up funds from the Boston University Department of Mechanical Engineering. We gratefully acknowledge this support. 

\section{Additional Information}
\label{sec:additional_info}

The entire Mechanical MNIST collection is available through the OpenBU Institutional Repository: \url{https://open.bu.edu/handle/2144/39371} \citep{MMequi,MMuniaxial}. 
The data used specifically for this paper is available at: \url{https://hdl.handle.net/2144/41357} \citep{MMmultifi}.
All code used to generate the data and metamodels presented in this paper is available through GitHub: \url{https://github.com/elejeune11/Mechanical-MNIST-Transfer-Learning}. 

\appendix

{\color{myColor}
\section{Additional datasets}
\label{apx:other_data}

\begin{table}[h]
{\footnotesize
 {\color{myColor}
\begin{tabular}{cp{8cm}}
\hline
Name             & Description                                                                                                                            \\ \hline
UE-CM-7-quad & uniaxial extension, $7 \times 7 \times 2$ quadratic triangular elements \\ 
UE-CM-4-quad & uniaxial extension, $4 \times 4 \times 2$ quadratic triangular elements \\ \hline
\end{tabular}}}
\caption{\label{tab:datasets_apx}{\color{myColor} Summary of two additional datasets, both within Mechanical MNIST. Both datasets are freely available with access details provided in Section \ref{sec:additional_info}.}}
\end{table}

\begin{figure}[h]
\begin{center}
\includegraphics[width=.75\textwidth]{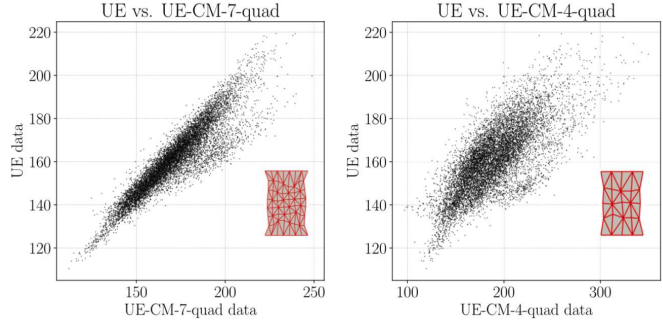}
\caption{\label{fig:apx_raw_data} {\color{myColor} Both plots illustrate the correlation between the standard Uniaxial Extension (UE) dataset and another dataset: UE-CM-7-quad and UE-CM-4-quad. All axis represent the QoI, change in strain energy $\Delta \psi$ at the end of the simulation, for each dataset. Each point corresponds to an input bitmap in the test set ($10,000$ points total per plot). }}
\end{center}
\end{figure}

\begin{figure}[h]
\begin{center}
\includegraphics[width=.75\textwidth]{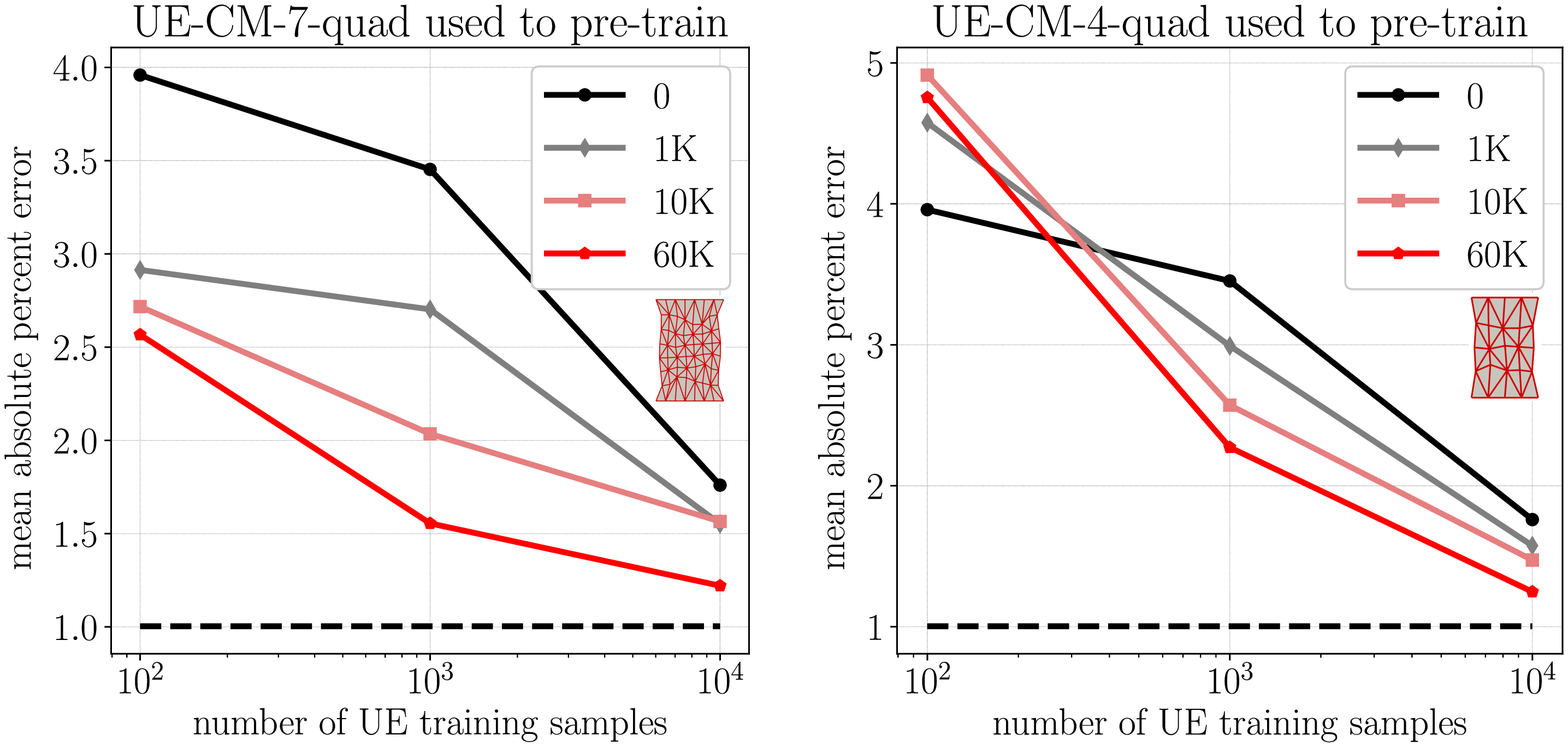}
\caption{\label{fig:apx_TL} {\color{myColor} These plots shows the test performance of pre-trained metamodels with respect to the number of training samples. In all plots, the black curve labeled ``$0$'' corresponds to metamodel performance without pre-training and the black dashed line corresponds to metamodel performance without pre-training for $60,000$ training points which represents the best performance achieved by the metamodel given the available data (see Section \ref{sec:mm_res}). The other curves, labeled ``$1$K'', ``$10$K'', and ``$60$K'', correspond to pre-trained metamodel performance where each metamodel is pre-trained with the indicated number of points from the related dataset. Left: UE prediction with UE-CM-7-quad used for pre-training; Right: UE prediction with UE-CM-4-quad used for pre-training.}}
\end{center}
\end{figure}

In Table \ref{tab:datasets_apx}, we describe two additional datasets: ``UE-CM-7-quad'' and ``UE-CM-4-quad.'' These datasets are identical to UE-CM-7 and UE-CM-4 respectively but with quadratic elements rather than linear elements. Details for accessing theses datasets are provided in Section \ref{sec:additional_info}. 
In Fig. \ref{fig:apx_raw_data}, we illustrate the relationship between each dataset and the UE dataset. These plots are consistent with Fig. \ref{fig:raw_data} in the main body of the text. Then, in Fig. \ref{fig:apx_TL}, we plot the results of predicting UE from metamodels pre-trained on each dataset. These plots are consistent with Fig. \ref{fig:plot_transfer_res} in the main body of the text.  

In Table \ref{tab:comp_cost}, we include these datasets in our evaluation of \textit{approximate} computational cost. 
When making choices about the type of simulations to run for pre-training, the key tradeoff is between the computational cost of generating the finite element simulation dataset, and the potential increase in metamodel performance. We note that, for this example, a finer discretization of linear elements provides a better tradeoff compared to a coarser discretization of quadratic elements. However, computational cost may vary with different choices of hardware and software, which may influence the balance of this tradeoff. Ultimately, the optimal parameters of the low fidelity model will be problem specific. 

\section{Additional information on dataset generation}
\label{apx:simulation_info}

Essential information about dataset generation is provided in Section \ref{sec:dataset} of this paper and in our previously published work \citep{Lejeune20a}. For reference, we use a compressible Neo-Hookean material model:
\begin{equation}
\ \psi =  \frac{1}{2} \mu  \bigg[ \textbf{F} : \textbf{F} - 3 - 2 \ln (\det \mathbf{F}) \bigg] + \frac{1}{2} \lambda \bigg[ \frac{1}{2} ((\det \mathbf{F})^2 - 1 ) - \ln (\det \mathbf{F}) \bigg]
\end{equation}
where $\psi$ is the strain energy, $\textbf{F}$ is the deformation gradient, and $\mu$ and $\lambda$ are the Lam{\'e} parameters equivalent to Poisson's ratio $\nu$ and Young's modulus $E$:
\begin{equation}
\ E = \frac{\mu \, ( \, 3 \lambda + 2 \mu \, ) \,}{ \lambda + \mu} \qquad \qquad \nu = \frac{\lambda}{2 \, ( \, \lambda + \mu \, )} \, . 
\end{equation}
We convert the MNIST bitmap images to material properties by dividing the material domain such that it corresponds with the grayscale bitmap and then we specify $E$ as 
\begin{equation}
\ E = \frac{c}{ 255.0} \, (100.0 - 1.0) + 1.0
\end{equation}
where $c$ is the corresponding value of the grayscale bitmap that can range from $0-255$. For all simulation, we keep Poisson's ratio fixed at $\nu = 0.3$ throughout the domain. 
Each simulation is thus a soft background material with ``digits'' that are two orders of magnitude stiffer than the background material. 
In Fig. \ref{fig:apx_raw_curves}, we show change in strain energy $\Delta \psi$ as a function of applied displacement for five randomly selected curves from each dataset. The same five input bitmaps are used to generate the curves in each plot. Consistent with the plots shown in Fig. \ref{fig:raw_data} and Fig. \ref{fig:apx_raw_data}, many of the datasets are both qualitatively and quantitatively similar. We note that all code used to generate dataset is freely available, with access details given in Section \ref{sec:additional_info}. 

\begin{figure}[p]
\begin{center}
\includegraphics[width=.9\textwidth]{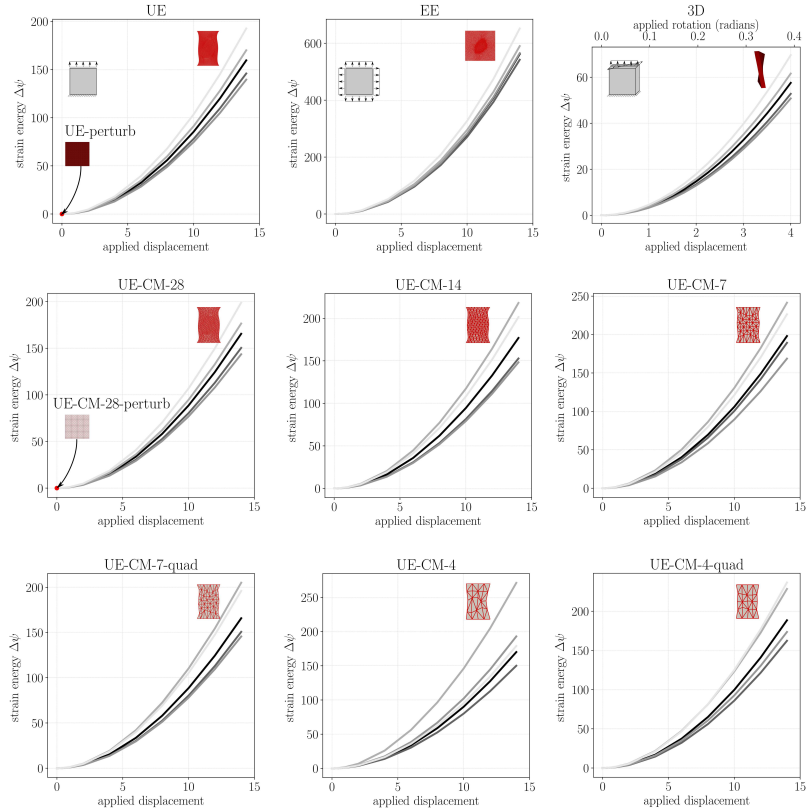}
\caption{\label{fig:apx_raw_curves} {\color{myColor} Five randomly selected change in strain energy $\Delta \psi$ vs. applied displacement curves from each dataset. Applied displacement units match the units of the $28 \times 28$ unit block. The same five input bitmaps are used for each plot. For the ``perturb'' datasets, the datapoint that corresponds to the perturbation simulation is shown in red. For all other datasets, $\Delta \psi$ is taken at the final step of applied displacement.}}
\end{center}
\end{figure}

\begin{table}[h]
\begin{center}
{\footnotesize
 {\color{myColor}
\begin{tabular}{cp{2cm}p{2cm}}
\hline
Name             &     \textit{Approximate} time for one simulation in seconds   &  \textit{Approximate} time for $60,000$ simulations in hours                \\ \hline
3D 		   &   $1,300$ s   & $30,000$ hrs  \\
UE               &       $400$ s  &    $6,600$ hrs     \\
UE-CM-28         &         $1.8$ s  &   $30$ hrs               \\
UE-CM-14         &         $1.4$ s   &   $23$ hrs                       \\
UE-CM-7          &          $1.4$ s    &  $23$ hrs                                       \\
UE-CM-4          &          $1.3$ s     & $22$ hrs                                              \\
UE-CM-7-quad          &        $2.8$ s  &  $46$ hrs                      \\
UE-CM-4-quad          &        $1.6$ s   &    $26$ hrs                                                   \\
UE-perturb       &                 $28$ s      &  $460$ hrs                                     \\
UE-CM-28-perturb &           $0.24$ s    & $4.1$ hrs                  \\ \hline
\end{tabular}}
\caption{\label{tab:comp_cost} {\color{myColor} Summary of the \textit{approximate} computational cost per simulation. Each simulation is run with open source software FEniCS \citep{Alnaes+etal15,Logg+etal12} at the Massachusetts Green High Performance Computing Center on a single core. All code to generate the data is freely available with access details provided in Section \ref{sec:additional_info}. We note that with different hardware or software choices computational time may vary significantly.}}}
\end{center}
\end{table}

In Table \ref{tab:comp_cost}, we list the \textit{approximate} computational cost of generating each dataset. With our chosen software and hardware, time for a single simulation ranges from $0.24$ seconds (UE-CM-28-perturb) to $1,300$ seconds (3D). In Fig. \ref{fig:plot_transfer_res}, we compare the performance of metamodels that are and are not pre-trained and argue that pre-training can lead to equivalent metamodel performance with a substantially cheaper to generate dataset. For example, in Fig. \ref{fig:plot_transfer_res}a, a pre-trained metamodel with $60,000$ UE-CM-28 simulations and $100$ UE simulations performs similarly to a metamodel trained with $60,000$ UE data points. The pre-training case requires approximately $41$ CPU hours whereas the case without pre-training requires approximately $6,600$ CPU hours. For the example shown in Fig. \ref{fig:plot_transfer_res}f, the pre-training case ($60,000$ UE-CM-28-perturb and $100$ UE) requires approximately $15$ CPU hours whereas the case without pre-training requires approximately $6,600$ CPU hours. From all examples shown in Fig. \ref{fig:plot_transfer_res}i, the most extreme example is UE-CM-28-perturb used to pre-train 3D. In this example, the pre-training case ($60,000$ UE-CM-28-perturb and $100$ 3D) requires approximately $40$ CPU hours, whereas the case without pre-training requires approximately $30,000$ CPU hours. Therefore, in our most extreme example, the metamodel built with transfer learning requires $0.13$\% of the CPU hours of the metamodel built without it. This clearly demonstrates the potential of transfer learning for enhancing metamodels of heterogeneous material behavior. 

\begin{table}[p]
{\scriptsize
 {\color{myColor}
\begin{tabular}{p{4cm}p{2cm}p{2.5cm}p{2.5cm}}
\hline
pre-trained on dataset-1 $\rightarrow$ additional training on dataset-2             &  number of pre-training points &  mean absolute test error on dataset-2 before transfer        &    mean absolute test error on dataset-2 after transfer with $100$ points additional training         \\ \hline
 UE-CM-28 $\rightarrow$ UE & $1,000$   & $6.08$\%  & $2.67$\% \\
           & $10,000$   &  $3.34$\% & $1.65$\%  \\
           & $60,000$   &  $3.16$\% & $1.06$\%  \\ \hline
 UE-CM-14 $\rightarrow$ UE & $1,000$  & $14.1$\% & $3.18$\% \\
            & $10,000$   & $7.98$\% & $2.46$\% \\
           & $60,000$   & $10.3$\%  & $2.00$\% \\ \hline
 UE-CM-7 $\rightarrow$ UE  & $1,000$ & $18.4$\% & $4.83$\% \\
            & $10,000$   & $22.6$\% & $4.23$\% \\
           & $60,000$   &  $24.1$\% & $4.42$\% \\ \hline
 UE-CM-4 $\rightarrow$ UE  & $1,000$ & $47.9$\% & $12.1$\% \\
            & $10,000$   & $47.7$\% & $12.2$\% \\
           & $60,000$   & $55.1$\% & $16.4$\% \\ \hline
 UE-perturb$^*$ $\rightarrow$ UE & $1,000$  & $37.1$\% & $2.79$\% \\
            & $10,000$   & $37.4$\% & $1.78$\% \\
           & $60,000$   & $38.3$\% & $1.39$\% \\ \hline
 UE-CM-28-perturb$^*$ $\rightarrow$ UE  & $1,000$ & $32.9$\% & $2.76$\% \\
            & $10,000$   & $34.9$\% & $1.77$\% \\
           & $60,000$   &  $34.6$\% & $1.34$\% \\ \hline
 UE-CM-7-quad $\rightarrow$ UE&  $1,000$  &  $10.9$\% & $2.91$\% \\
            & $10,000$   &  $8.32$\% & $2.72$\% \\
           & $60,000$   & $9.99$\% & $2.57$\% \\ \hline
 UE-CM-4-quad $\rightarrow$ UE &  $1,000$ & $16.6$\% & $4.58$\% \\ 
            & $10,000$   & $19.7$\% & $4.91$\% \\
           & $60,000$   & $18.1$\% & $4.75$\% \\  \hline
 EE $\rightarrow$ UE    & $1,000$ & $258$\%  & $3.44$\% \\
            & $10,000$   & $252$\% & $3.39$\% \\
           & $60,000$   &  $252$\% & $3.82$\% \\ \hline
 UE $\rightarrow$ EE    & $1,000$ & $71.0$\% & $2.23$\% \\
            & $10,000$   & $71.4$\% & $2.18$\% \\
           & $60,000$   & $71.7$\% & $1.79$\% \\ \hline
UE-CM-28-perturb$^*$ $\rightarrow$ 3D    & $1,000$ & $85.5$\% & $2.79$\% \\
            & $10,000$   & $80.2$\% & $2.05$\% \\
           & $60,000$   & $81.1$\% & $1.41$\% \\ \hline
\end{tabular}}}
\caption{\label{tab:err} {\color{myColor} This table accompanies Fig. \ref{fig:plot_transfer_res} and Fig. \ref{fig:apx_TL}. Here we compare the metamodel error on the dataset of interest (dataset 2) before and after using $100$ dataset-2 samples for additional training. $^*$As stated in the text, $\Delta \psi$ in the perturbation dataset is multiplied by $10^8$ before metamodel training to avoid numerical errors during the training process.}  }
\end{table}
\section{Metamodel performance before and after transfer}
\label{apx:fine}

In Fig. \ref{fig:plot_transfer_res} and Fig. \ref{fig:apx_TL}, we plot metamodel performance with respect to the number of ``dataset-2'' training samples (see schematic in Fig. \ref{fig:intro}). In each plot, each curve corresponds to a different number of ``dataset-1'' samples used for metamodel pre-training. In Fig. \ref{fig:plot_transfer_res} and Fig. \ref{fig:apx_TL}, there is a clear comparison between metamodels with pre-training (curves labeled $1$K, $10$K, and $60$K) and metamodels with no pre-training (black curve labeled $0$). We note that in most cases investigated, pre-training reduces metamodel error. An additional useful comparison is the metamodel performance without additional training on dataset-2 (i.e. transfer). If dataset-1 and dataset-2 are identical, the additional training will make little difference. In Table \ref{tab:err}, we compare mean absolute test error before transfer (column 3) and mean absolute test error after transfer with $100$ points of additional training (column 4). We note that the values in column 4 are datapoints in the Fig. \ref{fig:plot_transfer_res} and Fig. \ref{fig:apx_TL} plots. In all cases, the additional training decreases error when making predictions on dataset-2. In many cases, the metamodel without additional training would do a poor job making predictions on dataset-2.

}

\FloatBarrier
\newpage
\bibliography{mybibfile}

\end{document}